\documentclass[10pt, conference, letterpaper]{IEEEtran}
\IEEEoverridecommandlockouts

\usepackage{cite}
\usepackage{amsmath,amssymb,amsfonts,mathtools}
\usepackage{algorithmic}
\usepackage{graphicx}
\usepackage{textcomp}
\usepackage{xcolor}
\usepackage{bm}
\usepackage{mleftright}
\renewcommand{\hat}{\widehat}
\renewcommand{\left}{\mleft}
\renewcommand{\right}{\mright}

\usepackage[
  hidelinks,
]{hyperref}

\def\BibTeX{{\rm B\kern-.05em{\sc i\kern-.025em b}\kern-.08em
    T\kern-.1667em\lower.7ex\hbox{E}\kern-.125emX}}
\begin{document}

\title{
    Sparse Signal Recovery Using Log-Sum Regularization and Adaptive Smoothing%
    \thanks{A version of this work has been accepted for presentation at the 2026 International Symposium on Information Theory and Its Applications (ISITA2026).}
}

\author{\IEEEauthorblockN{Keisuke Morita}
\IEEEauthorblockA{\textit{Graduate School of Information Sciences,} \\
\textit{Tohoku University}\\
Sendai, Japan \\
mory22k@dc.tohoku.ac.jp}
\and
\IEEEauthorblockN{Masayuki Ohzeki}
\IEEEauthorblockA{\textit{Graduate School of Information Sciences,} \\
\textit{Tohoku University}\\
Sendai, Japan \\
mohzeki@tohoku.ac.jp}
}

\maketitle

\bstctlcite{IEEEexample:BSTcontrol}

\begin{abstract}
We study sparse signal recovery from noisy linear observations using nonconvex log-sum regularization.
The log-sum penalty reduces the shrinkage bias of $\ell_1$ regularization and more closely approximates the $\ell_0$ regularization, but its nonconvexity can make reconstruction algorithms unstable.
To mitigate this instability, we use an adaptive smoothing strategy that determines the smoothing parameter so that the scalar proximal operator remains continuous.
Using this proximal operator, we formulate the approximate message passing (AMP) algorithm and derive the corresponding state evolution (SE) recursion.
The fixed point of the SE recursion predicts the final mean squared error (MSE) and, in the noiseless limit, the exact-recovery phase transition.
To further investigate finite-dimensional reconstruction behavior, we implement an alternating direction method of multipliers (ADMM) algorithm.
In the noiseless setting, we find that the empirical success boundary of ADMM closely agrees with the SE-predicted phase transition.
In the noisy setting, we observe that AMP closely follows the SE prediction, whereas ADMM qualitatively reproduces the SE-predicted dependence of the final MSE on the regularization parameter.
A comparison with $\ell_1$ regularization shows that log-sum regularization is beneficial in low-density or high-measurement-rate regimes, whereas $\ell_1$ regularization remains preferable at higher densities and lower measurement rates.
\end{abstract}

\section{Introduction}

Sparse signal recovery aims to reconstruct a high-dimensional signal from incomplete or noisy observations by exploiting the fact that the signal has only a small number of nonzero components in a suitable basis.
Compressed sensing is a representative framework for this problem, in which sparse signals are reconstructed from a small number of linear measurements~\cite{Donoho2006-jx,Candes2006-he}.
This formulation arises in applications including astronomy~\cite{Wiaux2009-ut}, seismic data processing~\cite{Herrmann2008-eq}, wireless communications~\cite{Qin2018-ja}, and medical image processing~\cite{Lustig2007-zw}.

Let $\bm{x}^0 \in \mathbb{R}^N$ be the true signal.
In a standard noisy linear observation model, we observe
\begin{align}
\bm{y}
=
\bm{A}\bm{x}^0
+
\bm{w},
\end{align}
where $\bm{A} \in \mathbb{R}^{M \times N}$ denotes the measurement matrix and $\bm{w} \in \mathbb{R}^M$ denotes the noise vector.
The goal is to recover $\bm{x}^0$ from $\bm{y}$ and $\bm{A}$.
A standard approach is regularized least-squares reconstruction:
\begin{align}
\min_{\bm{x} \in \mathbb{R}^N}
~
\frac{1}{2}
\left\|
\bm{A}\bm{x}
-
\bm{y}
\right\|_2^2
+
\lambda_\text{pen}
\sum_{i=1}^N
R_i(x_i; \varepsilon),
\label{eq:objective_function}
\end{align}
where $\lambda_\text{pen} > 0$ is the regularization parameter and $R_i(\cdot; \varepsilon)$ is the penalty applied to the $i$-th component, with shape parameter $\varepsilon$ when applicable.
For the cases without noise or with small noise, the ideal sparsity penalty is $\ell_0$ regularization, $R_i(x_i) = \mathbb{I}_{\{x_i \neq 0\}}$, which directly penalizes the support size.
Since $\ell_0$-regularized reconstruction is computationally difficult, the least absolute shrinkage and selection operator (LASSO), corresponding to $\ell_1$ regularization $R_i(x_i) = |x_i|$, is a standard tractable alternative~\cite{Tibshirani1996-cx,Candes2006-he}.
Its typical performance and phase-transition behavior in compressed sensing have also been studied using statistical-mechanics methods~\cite{Rangan2009-el,Kabashima2009-gx,Ganguli2010-ef,Krzakala2012-is}.
A well-known limitation of LASSO, however, is its constant shrinkage, which can lead to overshrinkage of nonzero components of signals and production of false positives, especially when the number of measurements is limited~\cite{Fan2001-pc,Zou2006-qw}.

\begin{figure}[t]
\includegraphics[width=1.0\linewidth]{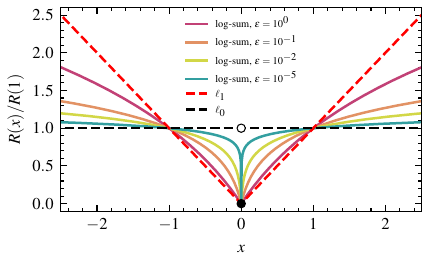}
\caption{
    The normalized LSP $R(x; \varepsilon) / R(1; \varepsilon)$ as a function of $x$ for various values of $\varepsilon$.
    The $\ell_1$ norm $R(x) = |x|$ and the $\ell_0$ pseudo-norm $R(x) = \mathbb{I}_{\{x \neq 0\}}$ are also shown for comparison.
}
\label{fig:logsum_penalty}
\end{figure}

Nonconvex penalties have been studied as alternatives that more closely approximate $\ell_0$ regularization while reducing the shrinkage bias of $\ell_1$ regularization.
Representative examples include $\ell_p$ ($0 < p < 1$) quasi-norm~\cite{Chartrand2007-so,Chartrand2008-rl}, log-sum penalty (LSP)~\cite{Candes2008-de,Shen2013-ed,Prater-Bennette2022-rz}, \textit{smoothly clipped absolute deviation} (SCAD)~\cite{Fan2001-pc}, and \textit{minimax concave penalty} (MCP)~\cite{Zhang2010-gb}.
More structured and optimization-oriented nonconvex penalties have also been proposed recently \cite{Sasaki2024-zd, Furuhashi2026-qr}.
Previous work has shown that such penalties can reduce bias and enlarge exact-reconstruction regions in compressed sensing~\cite{Fan2001-pc,Zou2006-qw,Chartrand2007-so,Chartrand2008-om,Foucart2009-je}.
Because nonconvex penalties make practical optimization sensitive to both the penalty shape and the algorithmic dynamics, several algorithmic strategies based on nonconvexity control and related message-passing methods have been proposed and analyzed~\cite{Sakata2018-ey,Sakata2021-du,Gu2025-wr}.

Among these nonconvex penalties, we focus on log-sum regularization with the penalty function
\begin{align}
R_i(x_i; \varepsilon) = \log\left( \frac{|x_i|}{\varepsilon} + 1 \right),
\end{align}
where $\varepsilon>0$ controls the smoothness of the penalty.
As shown in Fig.~\ref{fig:logsum_penalty}, its normalized form interpolates between $\ell_0$ and $\ell_1$: for fixed $x_i$, $R_i(x_i; \varepsilon) / R_i(1; \varepsilon) \to \mathbb{I}_{\{x_i \neq 0\}}$ as $\varepsilon \to 0^+$ and $R_i(x_i; \varepsilon) / R_i(1; \varepsilon) \to |x_i|$ as $\varepsilon \to \infty$.
In addition, like SCAD and MCP, the LSP is weakly convex and consequently has a parameter region in which its proximal operator remains continuous~\cite{Fan2001-pc,Zhang2010-gb,Prater-Bennette2022-rz}.
This property motivates an adaptive smoothing strategy that preserves strong sparsity promotion while maintaining algorithmic stability~\cite{Morita2026-fd}.

The main contributions of this work are as follows.
First, building on the adaptive smoothing and phase-transition analysis for log-sum minimization in the noiseless case developed in \cite{Morita2026-fd}, we extend the AMP algorithm using the scalar proximal operator of the LSP to the noisy case and give the SE prediction of its dynamics in the $N \to \infty$ limit.
Second, we implement the alternating direction method of multipliers (ADMM) for log-sum regularization and demonstrate that its minimum mean squared error (MSE) and qualitative dependence on $\lambda_\text{pen}$ are consistent with the SE prediction.
Third, we show that log-sum regularization can outperform LASSO primarily at low signal densities and high measurement rates, although LASSO remains preferable elsewhere.

\section{Preliminaries}
\label{sec:preliminaries}

\subsection{Problem Setting}

We consider the following synthetic setting.
Each entry of $\bm{A}$ is independently generated from a Gaussian distribution with mean $0$ and variance $1/N$, and the noise vector $\bm{w}$ is generated from a Gaussian distribution with covariance $\sigma^2 \bm{I}$.
We assume that the true signal $\bm{x}^0$ is sparse and has standard-normal nonzero components:
\begin{align}
P_\text{true}\left( \bm{x}^0 \right)
=
\prod_{i=1}^N
\left(
    \left(1-\rho\right)\delta(x_i^0)
    +
    \rho \phi(x_i^0)
\right),
\end{align}
where $\rho \in \left[0,1\right]$ is the signal density, $\delta(\cdot)$ is the Dirac delta function, and $\phi(\cdot)$ is the density of the standard normal distribution.
The expected number of nonzero components is thus $K = \rho N$.
Hereafter, we denote $\alpha = M / N$.

\subsection{Adaptive Smoothing of Log-Sum Penalty}

Unlike $\ell_1$ regularization, log-sum regularization is nonconvex, and algorithmic stability is therefore a practical concern.
We use \textit{adaptive smoothing} to address this challenge~\cite{Morita2026-fd}.
To describe this method, we briefly review the behavior of the proximal operator of the LSP.
For $R_i(x_i; \varepsilon)$, the scalar proximal operator used in AMP and ADMM is
\begin{align}
\operatorname*{prox}_{\lambda_\text{prox} R(\cdot; \varepsilon)}(h)
\coloneqq
\arg\min_{x \in \mathbb{R}}
\left\{
\frac{1}{2}\left(x - h\right)^2
+
\lambda_\text{prox} R(x; \varepsilon)
\right\},
\end{align}
where $\lambda_\text{prox} > 0$ is a positive parameter.
The closed-form analysis in \cite{Prater-Bennette2022-rz} has shown that its qualitative behavior is controlled by the relationship between $\varepsilon$ and $\lambda_\text{prox}$.
If $\varepsilon > \sqrt{\lambda_\text{prox}}$, the proximal objective is unimodal and the operator is continuous.
In this convex regime, the operator is given by
\begin{align}
\operatorname*{prox}_{\lambda_\text{prox} R(\cdot; \varepsilon)}(h)
=
\begin{cases}
    \operatorname{sign}(h) r_+(|h|) & \text{if } |h| > \lambda_\text{prox} / \varepsilon, \\
    0 & \text{otherwise},
\end{cases}
\end{align}
where $r_+(|h|) = (|h| - \varepsilon + \sqrt{(|h| + \varepsilon)^2 - 4\lambda_\text{prox}}) / 2$.
If $\varepsilon \le \sqrt{\lambda_\text{prox}}$, the objective can have two local minima and the proximal operator becomes discontinuous at the switching threshold $|h| = h_\text{th} \in [2\sqrt{\lambda_\text{prox}} - \varepsilon, \lambda_\text{prox} / \varepsilon]$.

Adaptive smoothing chooses the smoothing parameter $\varepsilon$ according to the proximal parameter $\lambda_\text{prox}$ by
\begin{align}
    \varepsilon = \sqrt{\lambda_\text{prox}} + \Delta_\varepsilon,
\end{align}
where $\Delta_\varepsilon$ is a constant parameter.
Setting $\Delta_\varepsilon \ge 0$ keeps the proximal operator continuous while retaining strong sparsity-promoting behavior.
For all numerical experiments in our study, we set $\Delta_\varepsilon = 10^{-10}$ consistently.
Because $\varepsilon$ changes with $\lambda_\text{prox}$ during the algorithm, \eqref{eq:objective_function} is not solved with a single fixed $\varepsilon$.
As a result, static predictions based on the replica method~\cite{Mezard1987-ck,Rangan2009-el,Kabashima2009-gx} cannot directly describe the algorithmic dynamics induced by adaptive smoothing.

\subsection{Approximate Message Passing and State Evolution}
\label{sec:amp_se}

AMP is an iterative algorithm that computes the expectation of the marginal posterior distribution under the formulation of compressed sensing as a Bayesian estimation problem~\cite{Donoho2010-mz}.
AMP iterates as follows:
\begin{align}
    \bm{h}^{[t]}
        &=
        \hat{\bm{x}}^{[t]}
        +
        \frac{N}{M}
        \bm{A}^\top \bm{z}^{[t]},
        \label{eq:amp_h}
    \\
    k^{[t]}
        &=
        \frac{1}{\alpha}
        \overline{
            \nabla S\left(
            \bm{h}^{[t]};
            \frac{\chi^{[t]} + \lambda_\text{pen}}{\alpha} R(\cdot; \varepsilon^{[t]})
            \right)
        },
        \label{eq:amp_k}
    \\
    \hat{\bm{x}}^{[t+1]}
        &=
        S\left(
        \bm{h}^{[t]};
        \frac{\chi^{[t]} + \lambda_\text{pen}}{\alpha} R(\cdot; \varepsilon^{[t]})
        \right),
        \label{eq:amp_x}
    \\
    \bm{z}^{[t+1]}
        &=
        \bm{y}
        -
        \bm{A} \hat{\bm{x}}^{[t+1]}
        +
        \bm{z}^{[t]} k^{[t]},
        \label{eq:amp_z}
    \\
    \chi^{[t+1]}
        &=
        (\chi^{[t]} + \lambda_\text{pen}) k^{[t]}.
    \label{eq:amp_chi}
\end{align}
Here, $\hat{\bm{x}}^{[t]}$ is the estimator, $\bm{z}^{[t]}$ is the corrected residual, and $\chi^{[t]}$ represents the susceptibility at iteration $t$.
$S$ is the componentwise proximal operator:
\begin{align}
S(h_i; \lambda R(\cdot; \varepsilon))
=
\operatorname*{prox}_{\lambda R(\cdot; \varepsilon)}(h_i).
\end{align}
The overline denotes the arithmetic mean over components: $\overline{\nabla S(\bm{h}; \cdot)} \coloneqq \sum_i S'(h_i; \cdot) / N$.

The last term in \eqref{eq:amp_z} is called the Onsager correction, which cancels the leading correlation induced by repeatedly using the same measurement matrix.
This cancellation allows the asymptotic dynamics of AMP with an i.i.d. Gaussian measurement matrix to be characterized by state evolution (SE)~\cite{Donoho2010-kk, Bayati2011-de}.
Let the empirical mean squared error (MSE) of AMP at iteration $t$ be
\begin{align}
    E^{[t]}
    =
    \frac{1}{N}
    \left\|
    \hat{\bm{x}}^{[t]}
    -
    \bm{x}^0
    \right\|_2^2 .
\end{align}
In the $N,M,K \to \infty$ limit with $M/N \to \alpha = O(1)$ and $K/N \to \rho = O(1)$, the corresponding large-system dynamics are characterized by the following SE recursions:
\begin{align}
    h_*^{[t]}
        &= x^0 + \sqrt{\sigma^2 + \frac{E^{[t]}}{\alpha}} z,
    \label{eq:se_h}
    \\
    k_*^{[t]}
        &=
        \frac{1}{\alpha}
        \mathbb{E}_{x^0, z}
        \left[
            S'\left(
                h_*^{[t]};
                \frac{\chi^{[t]} + \lambda_\text{pen}}{\alpha} R(\cdot; \varepsilon^{[t]})
            \right)
        \right],
    \label{eq:se_k}
    \\
    E^{[t+1]}
        &=
        \mathbb{E}_{x^0, z}
        \left[
            \left(
                S\left(
                    h_*^{[t]};
                    \frac{\chi^{[t]} + \lambda_\text{pen}}{\alpha} R(\cdot; \varepsilon^{[t]})
                \right)-
                x^0
            \right)^2
        \right],
    \label{eq:se_mse}
    \\
    \chi^{[t+1]}
        &=
        (\chi^{[t]} + \lambda_\text{pen})
        k_*^{[t]}.
    \label{eq:se_chi}
\end{align}
Here, the expectation is over the true-signal distribution of $x^0$ and an independent standard Gaussian variable $z$.

A fixed point of \eqref{eq:se_h}--\eqref{eq:se_chi} predicts the final MSE.
Because the LSP is nonconvex, multiple fixed points may exist in some parameter regimes~\cite{Morita2026-fd}.
The actual algorithm can therefore converge to different fixed points depending on the initialization and iteration history.
In the experiments, we used the AMP initialization $\hat{\bm{x}}^{[0]}=\bm{0}$ and $\chi^{[0]}=1$, and the corresponding SE initialization $E^{[0]}=\mathbb{E}[(x^0)^2]$ and $\chi^{[0]}=1$.

\subsection{Alternating Direction Method of Multipliers}

We also applied ADMM \cite{Boyd2010-hi} to LSP-regularized reconstruction using the same adaptive-smoothing principle as in AMP.
With the scaled dual variable $\bm{u}$ and the penalty parameter $\rho_{\text{ADMM}} > 0$, the ADMM updates are given by
\begin{align}
    \widetilde{\bm{x}}^{[t+1]}
    &=
        \left(
            \bm{A}^{\top}\bm{A}
            +
            \rho_{\text{ADMM}}\bm{I}
        \right)^{-1}
    \notag \\ & \qquad
        \left(
        \bm{A}^{\top}\bm{y}
        +
        \rho_{\text{ADMM}}
        \left(
            \bm{z}^{[t]}
            -
            \bm{u}^{[t]}
        \right)
        \right),
        \label{eq:admm_x}
        \\
    \hat{\bm{x}}^{[t+1]}
        &=
        \alpha_\text{relax}
        \widetilde{\bm{x}}^{[t+1]}
        +
        \left(1-\alpha_\text{relax}\right)
        \bm{z}^{[t]},
        \label{eq:admm_relax}
        \\
    \bm{z}^{[t+1]}
        &=
        S(\hat{\bm{x}}^{[t+1]} + \bm{u}^{[t]}; (\lambda_\text{pen}/\rho_{\text{ADMM}}) R(\cdot; \varepsilon)),
        \label{eq:admm_z}
        \\
    \bm{u}^{[t+1]}
        &=
        \bm{u}^{[t]}
        +
        \hat{\bm{x}}^{[t+1]}
        -
        \bm{z}^{[t+1]}.
        \label{eq:admm_u}
\end{align}
Here, the relaxation step \eqref{eq:admm_relax} is introduced to improve the numerical stability of the ADMM iterations.
The relaxation parameter is fixed to $\alpha_\text{relax} = 0.5$ throughout this study.
This value was empirically determined, and we found that moderate variations around $\alpha_\text{relax} = 0.5$ had little effect on the final MSE.
We also set $\rho_\text{ADMM}=1$ after preliminary tests with $\rho_\text{ADMM} \in \{10^{-1}, 1, 10^1\}$, among which $\rho_\text{ADMM}=1$ showed the most stable behavior.

In ADMM, $\varepsilon$ is fixed as $\varepsilon = \sqrt{\lambda_\text{pen}/\rho_{\text{ADMM}}} + \Delta_\varepsilon$ in the $\bm{z}$ update, whereas in AMP it changes at each iteration as $\varepsilon^{[t]} = \sqrt{(\chi^{[t]} + \lambda_\text{pen})/\alpha} + \Delta_\varepsilon$ in the $\hat{\bm{x}}$ update.
In this way, fixing $\rho_\text{ADMM}$ also fixes $\varepsilon$ and can affect the final MSE.
The ADMM experiments are therefore not intended to reproduce the AMP dynamics, but rather to examine whether practical finite-dimensional optimization with the LSP yields reconstruction performance consistent with the SE prediction.

For the noiseless case $\sigma^2 = 0$, we impose $\bm{y} = \bm{A}\bm{x}$, corresponding to \eqref{eq:objective_function} with $\lambda_\text{pen} \to 0^+$, and use the following ADMM updates:
\begin{align}
    \hat{\bm{x}}^{[t+1]}
    &=
    \bm{A}^\top \left( \bm{A} \bm{A}^\top \right)^{-1} \bm{y}
    \notag \\ & \qquad
    +
    \left(
        \bm{I} - \bm{A}^\top \left( \bm{A} \bm{A}^\top \right)^{-1} \bm{A}
    \right)
    \left(
        \bm{z}^{[t]} - \bm{u}^{[t]}
    \right),
    \\
    \bm{z}^{[t+1]}
    &=
    S(\hat{\bm{x}}^{[t+1]} + \bm{u}^{[t]}; (1/\rho_{\text{ADMM}}^{[t]}) R(\cdot; \varepsilon^{[t]})),
    \\
    \bm{u}^{[t+1]}
    &=
    \bm{u}^{[t]}
    +
    \hat{\bm{x}}^{[t+1]}
    -
    \bm{z}^{[t+1]}.
\end{align}
We experimentally found that this noiseless algorithm could not be stabilized by introducing the relaxation step.
Instead, we scheduled the parameter $\rho_\text{ADMM}$ to maintain stability.
In the experiments, the value was initialized at $\rho_\text{ADMM}^{[0]}=1$ and updated as $\rho_\text{ADMM}^{[t+1]} = 1.01 \rho_\text{ADMM}^{[t]}$ at each iteration.
We do not claim that this schedule is optimal and leave its systematic study for future work.
Accordingly, adaptive smoothing uses $\varepsilon^{[t]} = \sqrt{1/\rho_\text{ADMM}^{[t]}} + \Delta_\varepsilon$ in the noiseless $\bm{z}$ update.
In the both noisy and noiseless cases, the variables were initialized as $\bm{z}^{[0]} = \bm{u}^{[0]} = \bm{0}$.

\section{Phase Transition in the Noiseless Case}
\label{sec:phase_transition}

\begin{figure}
\includegraphics[width=1.0\linewidth]{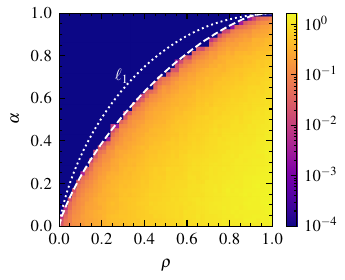}
\caption{
    Final MSE of ADMM reconstruction as a function of the measurement rate $\alpha$ and the signal density $\rho$ in the noiseless case $\sigma^2 = 0$.
    The ADMM results are averaged over $5$ independent trials with $N=10^3$.
    The white dashed line and the white dotted line represent the predicted phase transition for the log-sum and $\ell_1$ penalties, respectively.
}
\label{fig:phase_transition}
\end{figure}

We first analyzed the phase transition for reconstruction with log-sum regularization in the noiseless case $\sigma^2 = 0, \lambda_\text{pen} \to 0^+$.
$\bm{A}$, $\bm{x}^0$, and $\bm{w}$ were generated as described in Sec.~\ref{sec:preliminaries}.
Reconstruction performance was evaluated by the empirical MSE
\begin{align}
    \mathrm{MSE}
    =
    \frac{1}{N}
    \left\|
    \hat{\bm{x}}
    -
    \bm{x}^0
    \right\|_2^2 ,
\end{align}
where $\hat{\bm{x}}$ is the algorithmic reconstruction.
In this setting, the $(\alpha, \rho)$ space for compressed sensing recovery in the $N \to \infty$ limit is separated by a threshold $\alpha = \alpha_c(\rho)$ into a region where the method under consideration typically achieves exact recovery, $\mathrm{MSE}\approx 0$, and a region where it does not, $\mathrm{MSE}>0$.

Fig.~\ref{fig:phase_transition} compares the ADMM experimental results with the SE prediction.
ADMM experiments were run on $50 \times 50$ grids of $(\alpha, \rho)$ values in the ranges $\alpha \in (0, 1]$ and $\rho \in (0, 1]$.
In the SE computation, for each $\rho \in (0, 1]$, the smallest $\alpha$ for which the MSE fell below $10^{-4}$ was obtained by binary search.
Both ADMM and SE were run until the MSE fell below $10^{-4}$ or the number of iterations reached $10^3$.
For comparison, the typical reconstruction threshold of $\ell_1$ reconstruction \cite{Kabashima2009-gx} is shown as the white dotted line.
The empirically observed boundary is close to the SE prediction.
In addition, the threshold $\alpha_c(\rho)$ for LSP is lower than that for $\ell_1$ minimization, indicating that the log-sum version can achieve exact recovery with fewer measurements than $\ell_1$ minimization.

\section{Numerical Experiments in the Noisy Case}
\label{sec:numerical_experiments}

We performed experiments in the noisy case using a setting similar to the one in Sec.~\ref{sec:phase_transition}, but with $\sigma^2 = 10^{-2}$ and $\lambda_\text{pen} > 0$.
For both AMP and SE, we used damping with a damping factor of $0.2$ to improve stability, and the iteration was stopped when the MSE fell below $10^{-10}$ or the iteration count reached $10^3$.
For ADMM, the iteration was stopped when $\|\bm{z}^{[t+1]} - \hat{\bm{x}}^{[t+1]}\|_2 < 10^{-10}$ and $\rho_\text{ADMM} \|\bm{z}^{[t+1]} - \bm{z}^{[t]}\|_2 < 10^{-10}$ were satisfied or the iteration count reached $10^3$.

\begin{figure}[t]
\includegraphics[width=1.0\linewidth]{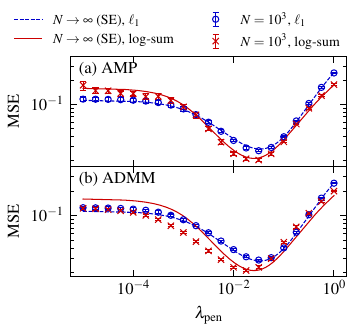}
\caption{
    Final MSE from the SE fixed-point prediction, together with (a) AMP and (b) ADMM results, as a function of $\lambda_{\mathrm{pen}}$.
    The measurement rate is $\alpha=0.9$, and the signal density is $\rho=0.4$.
    The AMP and ADMM symbols are averaged over $10$ independent trials with $N=10^3$.
}
\label{fig:mse_vs_lambda}
\end{figure}

First, we compared the SE-predicted final MSE with the algorithmic final MSEs.
Fig.~\ref{fig:mse_vs_lambda} compares the final MSE as a function of the regularization coefficient $\lambda_{\mathrm{pen}}$ for the SE fixed-point prediction, AMP experiments, and ADMM experiments.
For both $\ell_1$ and log-sum regularization, the MSE exhibits a U-shaped dependence on the regularization parameter $\lambda_{\mathrm{pen}}$.
The AMP results agreed well with the SE prediction over a wide range of $\lambda_{\mathrm{pen}}$.
Although the ADMM results did not fully agree with the SE prediction over the entire $\lambda_{\mathrm{pen}}$ range, the ADMM minimum MSE, $(2.1 \pm 0.1) \times 10^{-2}$ where the uncertainty denotes the standard error over 10 independent trials, agreed well with the SE prediction ($2.1 \times 10^{-2}$).

\begin{figure*}[t]
\includegraphics[width=1.0\linewidth]{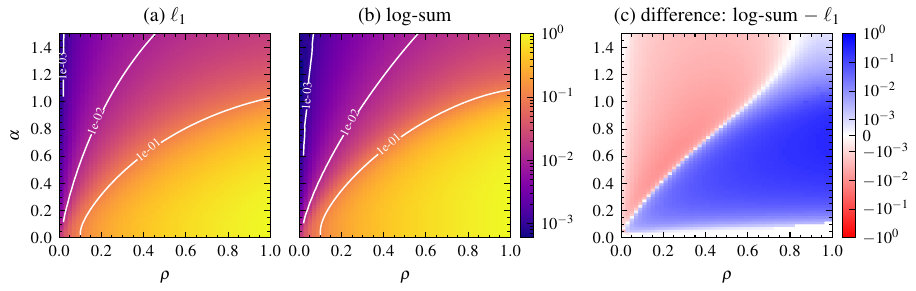}
\caption{
    Best MSE predicted by SE over $\lambda_{\mathrm{pen}}\in[10^{-4},10^2]$ for (a) $\ell_1$ and (b) log-sum regularization, and (c) the difference
    $d=\mathrm{MSE}_{\mathrm{logsum}}-\mathrm{MSE}_{\ell_1}$.
    The noise variance is $\sigma^2=10^{-2}$.
}
\label{fig:best_mse}
\end{figure*}

Next, under the same noise setting, we compared the minimum final MSEs of log-sum and $\ell_1$ regularization using the SE fixed-point prediction.
Fig.~\ref{fig:best_mse} plots the best final MSE predicted by SE in the $\rho$-$\alpha$ plane for $\alpha \in (0, 1.5]$ and $\rho \in (0, 1]$.
The best final MSE was obtained by searching $\lambda_\text{pen} \in [10^{-4},10^{2}]$.
For both penalties, the MSE decreases with increasing measurement rate $\alpha$ and decreasing signal density $\rho$.
Fig.~\ref{fig:best_mse}(c) presents the difference $d$ in the best final MSE between the log-sum and $\ell_1$ penalties on a symmetric logarithmic scale.
In this plot, the region $|d| \ge 10^{-3}$ is displayed on a log scale, and the other region is shown on a linear scale.
The blue region with positive values corresponds to lower MSE for the $\ell_1$ penalty, and the red region with negative values corresponds to lower MSE for the LSP.
This comparison shows that the relative performance of the two penalties depends on $(\rho,\alpha)$.
The LSP tended to yield a lower best final MSE mainly in the low-$\rho$ or high-$\alpha$ region.
In contrast, the $\ell_1$ penalty tended to yield a lower MSE in the high-$\rho$ and low-$\alpha$ region.

\section{Discussion and Future Work}

Our results indicate that SE predicts the noiseless phase boundary observed in the ADMM experiments under the parameter schedule used here and provides useful predictions for log-sum reconstruction performance.
They also suggest that log-sum regularization can outperform LASSO mainly when the signal density $\rho$ is low, or the measurement rate $\alpha$ is high.
When $\rho$ is high and $\alpha$ is low, on the other hand, LASSO can yield lower MSE.
One possible explanation is that the shrinkage bias induced by $\ell_1$ regularization to large components may suppress noise-induced amplitude growth, thereby unintentionally stabilizing the reconstruction.

Future work should include a more detailed analysis of the dependence on noise variance and finite-size effects, as well as a systematic study of the ADMM parameter schedule.
It would also be worth exploring the combination of the LSP with $\ell_2$ regularization, which may provide robustness similar to that of the elastic net~\cite{Zou2005-gg} while retaining nonconvex sparsity promotion.

\section{Acknowledgment}

This work was supported by programs for bridging the gap between R\&D and IDeal society (Society 5.0) and Generating Economic and social value (BRIDGE) and Cross-ministerial Strategic Innovation Promotion Program (SIP) from the Cabinet Office (No. 23836436).

\bibliographystyle{IEEEtran}
\bibliography{main}

\end{document}